\documentclass[useAMS,usenatbib,aas_macros]{mn2e}
\usepackage{aas_macros}

\newcommand{\equ}[1]{eq.~(\ref{eq:#1})}
\newcommand{\equs}[1]{eqs.~(\ref{eq:#1})}

\newcommand{\se}[1]{\S\ref{sec:#1}}
\newcommand{\fig}[1]{Fig.~\ref{fig:#1}}

\newcommand{\Fig}[1]{Figure~\ref{fig:#1}}
\newcommand{\be}{\begin{equation}}
\newcommand{\ee}{\end{equation}}

\newcommand{\msun}{M_\odot}

\newcommand{\lsun}{L_\odot}
\newcommand{\ifm}[1]{\relax\ifmmode#1\else$\mathsurround=0pt #1$\fi}
\newcommand{\kms}{\ifmmode\,{\rm km}\,{\rm s}^{-1}\else km$\,$s$^{-1}$\fi}

\newcommand{\kpc}{\,{\rm kpc}}

\newcommand{\ltsima}{$\; \buildrel < \over \sim \;$}
\newcommand{\lsim}{\lower.5ex\hbox{\ltsima}}
\newcommand{\gtsima}{$\; \buildrel > \over \sim \;$}
\newcommand{\gsim}{\lower.5ex\hbox{\gtsima}}
\newcommand{\prop}{\propto}

\newcommand{\vv}{{v}}
\newcommand{\rr}{{r}}
\newcommand{\mm}{{m}}
\newcommand{\Vd}{V_{\rm d}}
\newcommand{\Rd}{R_{\rm d}}

\newcommand{\Ve}{V_{\rm e}}
\def\Re{R_{\rm e}}

\newcommand{\ms}{{\cal M}}
\newcommand{\mse}{{\cal M}_{\rm e}}
\newcommand{\msd}{{\cal M}_{\rm d}}
\newcommand{\mb}{M_{\rm bar}}

\newcommand{\taus}{{\tau_{*}}}

\newcommand{\taud}{\tau_{\rm d}}
\newcommand{\alphad}{\alpha_{\rm d}}
\newcommand{\betad}{\beta_{\rm d}}
\newcommand{\taue}{\tau_{\rm e}}
\newcommand{\alphae}{\alpha_{\rm e}}
\newcommand{\betae}{\beta_{\rm e}}
\newcommand{\ttau}{\tilde\tau}
\newcommand{\talpha}{\tilde\alpha}
\newcommand{\tbeta}{\tilde\beta}

\newcommand{\ttaue}{\tilde\tau_{\rm e}}
\newcommand{\talphae}{\tilde\alpha_{\rm e}}
\newcommand{\tbetae}{\tilde\beta_{\rm e}}

\newcommand{\gamd}{\gamma}
\newcommand{\deld}{\delta_{\rm d}}

\newcommand{\Mv}{M_{\rm v}}
\newcommand{\Rv}{R_{\rm v}}
\newcommand{\Vv}{V_{\rm v}}
\newcommand{\Vm}{V_{\rm max}}
\newcommand{\tauv}{{\tau_{\rm v}}}
\newcommand{\cv}{c_{\rm v}}

\title[Progenitors of Ellipticals]
{The Dissipative Merger Progenitors of Elliptical Galaxies}

\author[A. Dekel \& T.J. Cox]
{Avishai Dekel$^{1}$ \& Thomas J. Cox$^{2}$\\
\\
$^1$Racah Institute of Physics, The Hebrew University, Jerusalem 91904 Israel\\
$^2$Center for Astrophysics, Harvard University, 60 Garden Street, Cambridge
MA 02138, USA\\
dekel@phys.huji.ac.il; tcox@cfa.harvard.edu}

\begin{document}

\pagerange{\pageref{firstpage}--\pageref{lastpage}} \pubyear{2002}

\maketitle

\label{firstpage}

\begin{abstract}
We address the deviations of the scaling relations of elliptical galaxies from 
the expectations based on the virial theorem and homology, including the 
``tilt" of the ``fundamental plane" and the steep decline of density with 
mass.  We show that such tilts result from dissipative major mergers once the 
gas fraction available for dissipation declines with progenitor mass, and 
derive the scaling properties of the progenitors.  We use hydrodynamical 
simulations to quantify the effects of 
major mergers with different gas fractions on the structural properties of 
galaxies.  The tilts are driven by the differential shrinkage of the effective 
stellar radius as a function of dissipation in the merger, while the correlated
smaller enhancements in internal velocity and stellar mass keep the slope of 
the velocity-stellar mass relation near $V\prop \ms^{1/4}$.  The progenitors 
match a straightforward model of disc formation in $\Lambda$CDM haloes.  Their 
total-to-stellar mass ratio within the effective radius varies as 
$M/\ms \prop \ms^{-(0.2-0.3)}$, consistent with the effect of supernova 
feedback.  They are nearly homologous in the sense that the dark-matter 
fractions within the effective and virial radii scale with mass in a similar 
way, with only a weak decline of density with mass.  The progenitor radius is 
roughly $R \prop \ms^{0.3}$, compatible with today's intermediate late-type 
galaxies.  This may indicate that the latest dissipative mergers in the history
of ellipticals involved relatively big discs.  The dissipative gas-to-stellar 
mass ratio is predicted to decline $\prop \ms^{-(0.4-0.6)}$, similar to
the observed trend in today's blue-sequence galaxies. Such a trend should be 
observed in relatively massive gaseous galaxies at $z \sim 1-4$.  The 
corresponding ``baryonic" relations are consistent with homology and spherical
virial equilibrium, $V \propto R \propto \mb^{1/3}$.
\end{abstract}


%%%%%%%%%%%%%%%%%%%%

\section{Introduction}
\label{sec:intro}

The global properties of elliptical galaxies show non-trivial correlations
whose origin is one of the most interesting issues in the study of galaxy
formation. These correlations may tell us whether the ellipticals 
could have indeed formed by mergers, and if so what were the
properties of their initial progenitors.
We address these issues in this paper.

% correlations
We appeal here to 
three observable global properties of spheroidal stellar systems: 
the total luminosity $L$, the ``effective" radius $R$ encompassing
half the total luminosity in projection,
and the effective velocity dispersion $\sigma$.\footnote{In the literature,
the radius is
sometimes replaced by surface brightness. The use of ``effective" surface 
brightness $I=L/R^2$ is equivalent to using $R$, but not exactly so when $I$
and $\sigma$ refer instead to the ``central" quantities. 
Nevertheless, as long as the density profile is close to universal, 
the central quantities should roughly scale with
the effective quantities.}
The galaxies lie on a ``Fundamental Plane" (FP) in the three-dimensional
parameter space defined by the logarithms of these quantities 
\citep{dressler87,djorgovski87,bernardi03_III},  
with very little scatter about it.
The projections of this plane onto the planes defined by the logarithms
of the axes $\sigma$-$L$ and $R$-$L$ are well fit by straight lines,
namely ``scaling relations", following \cite{faber76} and
\citet{kormendy77}. 

% tilts
As specified in the following section,
the fundamental plane is ``tilted" compared to the plane implied by the
virial theorem. While part of this tilt can be attributed 
to systematic variations
in stellar populations \citep{faber87}, the other part indicates a 
breakdown of the homology, or self-similarity, of the elliptical family, 
which can be expressed as a systematic increase of dark-matter fraction 
within the effective radius as a function of galaxy mass
\citep{padi04,onorbe05}.
Another indication of a breakdown in homology is provided by the
scaling relations, which imply a steep decline of mean effective
density with mass,
steeper than the moderate decline expected based on the properties of 
dark-matter haloes in the standard $\Lambda$CDM cosmology \citep{bullock01_c}. 
Supporting evidence for non-homology comes from the fact that 
the surface-brightness profiles of larger ellipticals appear to be
steeper, with a larger Sersic index \citep{caon93,graham03}.
We seek understanding of these deviations from homology.

% dissipative mergers. progenitors.
In this paper, we address the structural changes induced by dissipative
mergers.
Our simple analytic analysis is based on power-law recipes for
how the changes in the structure parameters due to mergers 
depend on the effective
gas fraction in the progenitors. We deduce these recipes from a suite of
disc merger simulations with a varying degree of gas fraction.
This analysis allows us to connect the observed FP and scaling relations
of ellipticals with the properties of their progenitors via simple equations. 
We use them to put constraints on the properties of these progenitors,
their scaling relations and especially the systematic trend of gas fraction 
with mass in them.

% outline
In \se{obs} we spell out the observed slopes of the scaling relations
and fundamental plane, and the implied deviations from homology.
In \se{tilt} we discuss the possible origin of these structural deviations,
and motivate the study of the role of dissipation in mergers.
In \se{equations} we formulate the merger process in terms of simple
power-law relations and derive equations that 
connect the observed relations for elliptical galaxies
with the properties of their original progenitors, including the
variation of gas fraction with mass.
In \se{sim} we use merger simulations to calibrate the necessary power-law
approximations for the dependence of the structural changes in mergers
on the gas fraction.
In \se{results} we use the recipes derived from the simulations to solve
for the properties of the progenitors given the observed structural relations.
In \se{interpret} we compare the required progenitor properties to what
might be expected based on other considerations.
In \se{conc} we summarize our conclusions.

%%%%%%%%%%%%%%%%%%%%%%%%%%%%%%%%%%%%%%
\section{Observed Scaling Relations}
\label{sec:obs}

% scaling
We write the observed scaling relations for elliptical galaxies as 
\be
\sigma \propto L^{\talpha} \ , \quad R \propto L^{\tbeta} \ ,
\label{eq:fpp0}
\ee
where $L$ is the absolute luminosity in a given band.
The slopes $\talpha$ and $\tbeta$ were determined by \citet{bernardi03_II}
for a large sample from the Sloan Digital Sky Survey (SDSS),
applying a maximum-likelihood analysis for the conditional average of the 
log of the relevant quantity at a given luminosity, namely
\be
\langle \log \sigma | L \rangle = \talpha \log L + const., \quad {\rm etc.}
\ee
Unless specified otherwise, we assume hereafter analogous definitions of 
slopes for all the power-law relations. 
The most recent values based on revised photometry in the $r$ band
(M. Bernardi, private communication) are
$\talpha=0.230 \pm 0.012$ and $\tbeta=0.704 \pm 0.025$,\footnote{revised from
the earlier published values of 0.255 and 0.632.}
with conditional standard deviations of the logs $0.063$ and $0.115$ 
respectively. The slopes in the other SDSS bands are not very different.

% FP
The Fundamental Plane as determined from SDSS by \citet{bernardi03_III}
is similar to the original FP \citep{dressler87,djorgovski87}
and can be approximated by 
\be
\sigma^2 R \propto L^{1+\ttau} \ ,
\label{eq:fp}
\ee
with $\ttau \simeq 0.17$ and a very small scatter about it.
Given the way the slopes are determined, the value of $\ttau$ is expected
to be related to the slopes of the scaling relations via
$1+\ttau = 2\talpha+\tbeta$.
The revised slopes of the scaling relations indeed imply $\ttau \simeq 0.164$.

% taus
For the purpose of studying structural and dynamical issues we rather
replace the luminosity $L$ with the corresponding stellar mass $\ms$ 
(which for ellipticals roughly constitutes the whole baryonic mass). 
We parameterize the systematic variation in stellar mass-to-light ratio
due to stellar populations by   
\be
\frac{\ms}{L} \prop L^{\taus} \ ,
\ee
and for the sake of simplicity ignore the scatter about this relation.
The structural relations for ellipticals within the effective radius become
\be
\sigma \propto \ms^{\alpha} \ , \quad 
R \propto \ms^{\beta} \ , \quad
\sigma^2 R \propto \ms^{1+\tau} \ ,
\label{eq:scaling_s}
\ee
with
\be
\alpha=\frac{\talpha}{1+\taus} \ , \quad
\beta=\frac{\tbeta}{1+\taus} \ , \quad
1+\tau=\frac{1+\ttau}{1+\taus} \ .
\label{eq:slopes_s}
\ee
The value of $\taus$ is believed to be on the order of $0.1$ 
\citep[e.g.][]{shen03}.
It can therefore make only a $\sim 10\%$ difference to the scaling
slopes $\alpha$ and $\beta$, but a major difference to the FP tilt $\tau$.

%Another analysis of SDSS data \citep{shen03} yields in the regression of 
%$R$ on %$M_*$ a slope of $\beta/(1+\taus) \simeq 0.56$, indicating 
%$\taus \simeq 0.12$.

% virial
The scaling relations are non-trivial in two ways.
First, the FP is tilted relative to the virial plane, 
\be
\sigma^2 R \propto M \ ,
\label{eq:virial}
\ee
which is expected to be roughly valid for the total mass within the 
effective radius if the ellipticals are a homologous family. 
The tilt reflects a systematic variation of the total mass-to-light ratio 
within the effective radius, $M/L \prop L^{\ttau}$. This could be partly 
due to stellar populations and partly a structural 
tilt,
\be
\frac{M}{\ms} \prop \ms^{\tau} \ ,
\quad \tau = 2\alpha+\beta-1 \ .
\label{eq:tilt}
\ee
If indeed $\tau>0$, it represents a breakdown of homology for these systems,
indicating that the mass fraction of dark matter within the effective 
radius is increasing with mass.
Furthermore, any non-negative tilt ($\tau \geq 0$) for the merger remnants
is non-trivial because the progenitors are actually expected to
show a negative tilt relative to the virial plane. 
This is if the variation of $M/\ms$ with progenitor mass is attributed 
to the decreasing effectiveness of supernova feedback in suppressing star
formation as a function of mass \citep{ds86,dw03}.

% density
Second,
the virial relation also implies that the mean total mass density within
the effective radius varies with $\ms$ as 
\be
\rho \prop \frac{\sigma^2}{R^2} \prop \ms^\delta\ , 
\quad \delta=2\alpha-2\beta \ .
\label{eq:rho}
\ee 
The observed values of $\talpha$ and $\tbeta$ correspond to $\delta \simeq
-0.95/(1+\taus)$, namely a rather steep decline of density with mass,
which provides an even stronger evidence for a homology breakdown.
The simplest self-similar model of simultaneous spherical collapse 
predicts virial equilibrium at a constant mean density,
with $\alpha=\beta=1/3$. Indeed, numerical studies of dark-matter haloes in 
$\Lambda$CDM cosmological simulations reveal only a weak systematic
density variation, weaker than $\delta \simeq -0.25$ 
\citep[based on][]{bullock01_c}.
The steep decline observed for ellipticals thus implies a breakdown of 
homology.  Unlike the tilt of the FP, the density decline cannot be 
explained by a variation in the stellar $\ms/L$ for any sensible value 
of $\taus$.

In our analysis below we adopt $\taus = 0.1$, but this particular choice 
within the range of uncertainty does not have a significant effect on our 
conclusions. The targets for our modeling thus become
$\alpha \simeq 0.21$ and $\beta \simeq 0.64$, with the associated
$\tau \simeq 0.06$ and $\delta =\simeq -0.86$.

%2 
%%%%%%%%%%%%%%%%%%%%%%%%%%%%%%%%%%%%%%%
\section{Possible Origin of the Tilts}
\label{sec:tilt}

%D10 e's by mergers
When addressing the possible physical origin of the structural tilts,
we adopt the standard paradigm of galaxy formation. It assumes that the 
accretion of cold gas in dark-matter haloes leads to central gaseous discy 
systems in which stars form \citep{wr78,fe80,mmw98}.  Major mergers between 
such systems of comparable masses 
produce more stars, consume the available cold gas, and transform the stellar 
systems into elliptical configurations 
\citep{fall79,jones79,aarseth80,barnes92}. 
In haloes more massive than $\sim 10^{12}\msun$, there is a shutdown of 
gas cooling and star formation,
allowing the stellar population to age passively and turn ``red and dead"
\citep[e.g.][]{bd03,binney04,keres05,db06,croton05,cattaneo06}.
These ellipticals may grow further by non-dissipative mergers with other 
ellipticals.  
One of the open questions is whether the dissipative mergers in the history
of big ellipticals involved big systems in a relatively late epoch or small
systems earlier on, followed by major non-dissipative mergers. 

% introduce growth factors
As a simple and general model for the mergers that transform the initial 
systems into ellipticals,
we assume that progenitor galaxies of type ``d" are transformed by
mergers to remnant galaxies of type ``e". Each galaxy is 
characterized by the three global quantities: 
a stellar mass $\ms$,
an effective stellar radius $R$, 
and a characteristic velocity $V$ (circular velocity for discs or
velocity dispersion for spheroids, up to a multiplicative factor of order 
unity).
The dimensionless {\it growth factors\,} 
describing the structural changes due to mergers, 
$\Re/\Rd$, $\Ve/\Vd$ and $\mse/\msd$,  
are determined by the merger dynamics. If the progenitors were completely 
homologous, and all the processes involved in a merger were scale free, 
then these factors were independent of mass, and the mergers would 
have led from power-law scaling relations to similar power laws,
though in general with different zero-points. 
This would have meant that the structural plane in $\ms$-$R$-$V$ 
and its projections are preserved under mergers, but perhaps with an 
increased scatter.
 
In particular, if the elliptical galaxies 
are homologous, and the merger dynamics involves gravitational dynamics
only, then the FP should be preserved under a few major 
mergers.\footnote{Phase
mixing and other gravitational relaxation processes may eventually
push such systems toward the virial plane obeyed by the total mass, 
but this is apparently a long
term process that requires many successive mergers, and, anyway, it does 
not produce a tilt toward the FP.} 
Such a behavior has indeed been noticed in N-body simulations of
dissipationless mergers \citep{boylan05,robertson06}. 
The question is thus {\it how did the elliptical galaxies make it to the FP 
in the first place}. This requires a breakdown of the self-similarity
of at least one property of the progenitors or the processes acting during
the mergers.
The dimensionless growth factors 
should depend on mass due to some important merger
characteristic that varies strongly with mass.

The gravitational process associated with spherical collapse and virial
equilibrium are not expected to break the self-similarity of the progenitors
in a major way. The systematic variation of the characteristic
collapse time with mass introduces a relatively weak variation in halo
concentration which is not enough for explaining the deviations from homology.
The origin of angular momentum via tidal torques is also largely 
self-similar \citep{fe80,bullock01_j}, implying disc sizes which are close
to self-similar.
The distribution of merger orbit parameters is not expected to vary
significantly with mass either \citep{vitvitska02}.

% contraction due to dissipation
On the other hand, gas dissipation may lead to a more substantial breakdown
of homology. 
Once the progenitors contain gas, the {\it dissipation\,} involved in the 
merger leads to a more centrally concentrated stellar system than in the 
non-dissipative case.  The tidal interactions during the close passages 
induce collisions and shocks which condense and heat the gas.  This is 
followed by enhanced {\it radiative cooling\,} and {\it star formation}, 
ending up with a more compact stellar system deeper in the potential well.

% other evidence for dissipation
Dissipation is indicated, for example, by the high phase-space density
in the centers of ellipticals \citep{hernquist93,robertson06},
and by the tendency for oblateness and kinematic misalignment of ellipticals
compared to simulated merger remnants \citep{cox06}.

% Robertson et al.
A property of the gas which depends on galaxy mass and could therefore produce
structural tilts in the scaling relations
is the {\it cooling efficiency\,} on a dynamical timescale, 
which declines with halo mass \citep[e.g.,][]{ro77,blum84}.  
A tilt of this sort has been seen in remnants of 
hydrodynamical merger simulations by \citet{robertson06}. 
Their Fig.~11 shows a tilt in the $R$-$\ms$ relation when the gas fraction is
0.4 (or higher), but the change from $\beta \simeq 0.44$ to $0.51$ is small
compared to the required tilt.
A careful inspection of this figure 
reveals that for a fixed gas fraction the tilt is noticeable only 
for giant ellipticals with stellar masses above $\sim 5\times 10^{11}\msun$,
while the observed tilts extend far below this mass.

% varying gas fraction
A more pronounced violation of the self-similarity of the dissipative
processes may arise from a systematic variation of the {\it gas fraction\,} 
with progenitor mass.  The relative amount of gas in the merging systems
has a strong effect on the resulting stellar mass and structure of the remnant.
More gas naturally results in more new stars, and the higher degree of
dissipation involved in the merger leads to an enhanced shrinkage in the
radius $R$ compared to the non-dissipative case. 
Based on a combination of physical considerations and numerical simulations
we have learned that the fraction of new stars and the radiative energy 
loss during the merger
are both roughly proportional to the effective gas fraction \citep{covington06}.
These two effects make the decline in $R/\ms$ due to mergers
depend strongly on the gas fraction.
The associated increase in velocity dispersion 
makes the ratio $V/\ms$ rather insensitive to the mergers and the gas fraction.
Together, the increase in $V$ is not enough to compensate for the strong
decrease in $R/\ms$, yielding a strong gas-dependent decline in the combination 
relevant to the FP, $V^2R/\ms$.
Thus, a proper decline of gas fraction 
with progenitor mass could, in principle, lead to the required tilts.

Such a gradient of gas fraction with progenitor mass is likely for 
several reasons.
We first note that today's disc galaxies along the blue sequence
show a strong trend which can be crudely fitted by a power law,
$g \propto \ms^{-\gamma}$,
where $g$ is the gas-to-star mass ratio and $\gamma \simeq 0.4$
\citep[][Figure 2c]{kannappan04}.
The earlier progenitors of elliptical galaxies
are expected in general to contain more gas, before it has been consumed
into stars or removed from the systems. 
Since small-mass galaxies tend to merge earlier than massive galaxies
in any given neighborhood, the major mergers involving small progenitors
are expected to be more dissipative than the mergers of massive progenitors.
The gas-fraction gradient may also be a result of the higher 
efficiency of quiescent star formation in larger discs, being determined by the 
higher surface density \citep{kennicutt89} 
rather than the three-dimensional density. 
If the mean 3D density is roughly independent of mass at any given epoch,
as implied by the naive spherical collapse model and tidal torque theory, 
the surface density is expected to be proportional to radius. 
Finally, a lower star formation rate at lower masses is expected to result 
from the more efficient supernova feedback slowing down the quiescent star 
formation preferentially in smaller galaxies \citep{ds86,dw03,db06}.
 
Hereafter, we take $g$ to refer to the gas that is effectively available for  
dissipation during the merger, i.e., combining the initial amount of gas 
and the dependence of the cooling efficiency on mass.

%3
%%%%%%%%%%%%%%%%%%%%%%%%%%%%
\section{Structural Tilts by Mergers}
%Pre-merger versus Post-merger Properties}
\label{sec:equations}

We assume that the progenitors (``d") and merger remnants (``e")
lie on (tilted) planes and scaling relations as in \equ{scaling_s}
in the corresponding parameter spaces $\Vd$-$\Rd$-$\msd$ and
$\Ve$-$\Re$-$\mse$ respectively, where all quantities refer to the spheres
defined by the effective radii encompassing half the stellar mass.
The six relations can be expressed in terms of the growth
factors $\Re/\Rd$, $\Ve/\Vd$ and $\mse/\msd$:
\be
\left(\frac{\Ve}{\Vd}\right)^2
\left(\frac{\Re}{\Rd}\right)
\prop \left(\frac{\mse}{\msd}\right)^{1+\taue} \msd^{\taue-\taud} \ ,
\label{eq:plane1}
\ee
and 
\be
\frac{(\Ve/\Vd)}{(\mse/\msd)^{\alphae}} \prop \msd^{\alphae-\alphad} \ ,
\quad
\frac{(\Re/\Rd)}{(\mse/\msd)^{\betae}} \prop \msd^{\betae-\betad} \ .
\label{eq:scaling1}
\ee

We then parameterize the structural growth factors in the mergers as 
functions of $g$ by the power laws
\be
\frac{\Re}{\Rd} \propto g^{\rr} 
\quad
\frac{\Ve}{\Vd} \propto g^{\vv}
\quad
\frac{\mse}{\msd} \propto g^{\mm} \ .
\label{eq:growth}
\ee
Finally, we assume that the mean dependence of $g$ on progenitor mass can 
be parameterized by,
\be
g \propto \msd^{-\gamd} \ .
\label{eq:gamma}
\ee

When substituting these power laws in \equ{plane1} and \equs{scaling1}, 
we obtain useful relations between the power indices involved,
\be
2\vv+\rr = (1+\taue)\mm - \gamd^{-1}(\taue-\taud) \ ,
\label{eq:tau}
\ee
\be
\gamd(-\vv +\mm\alphae) = (\alphae-\alphad) \ ,
\label{eq:v}
\ee
\be
\gamd(-\rr+\mm \betae) = (\betae-\betad) \ .
\label{eq:r}
\ee
These three equations are degenerate; only two are independent, with
$\tau=2\alpha+\beta-1$ for ``d" and for ``e".
Once we know the growth-factor indices $\rr$, $\vv$ and $\mm$ from theory,
these equations connect the final structural tilts to the properties of the 
progenitors, which are described by three independent variables:
the gas-fraction gradient index $\gamd$, and the slopes $\alphad$ and $\betad$
(one of which can be replaced by $\taud$ or $\deld$).
For given values of the observed parameters $\alphae$ 
and $\betae$ (or $\taue$),
the solutions for the progenitors are a one-parameter family.
Namely, for any assumed value of $\gamd$ there is a unique solution for
$\alphad$, $\betad$, etc.

%4
%%%%%%%%%%%%%%%%%%%%%%%%%%%%%%%%%%%%%%%%%%%%%%%%%
\section{Merger Simulations: Gas Dependence}
\label{sec:sim}

In order to determine the dependence of the structural growth factors
on the gas fraction [\equ{growth}],
we ran two sets of simulations of typical mergers of disc galaxies,
in which the initial conditions were identical except for the gas fraction 
in the disc; the initial gas-to-stars mass ratio $g$ ranges from 
$\sim 0.1$ to 3.
The two sets differ only in their bulge-to-disc ratio.

The fiducial case is G3-G3 from the suite of merger simulations 
in \citet{cox04}.
Two identical spiral galaxies are put on a parabolic orbit,
each consisting of stellar and gaseous discs and a stellar bulge,
all embedded in a $\Lambda$CDM halo.
The gravitational and hydrodynamical evolution is
followed using the entropy-conserving, gravitating,
Smoothed Particle Hydrodynamics (SPH) code 
GADGET \citep{springel01_gadget,springel02_entropy}.
Gas cooling, star formation and supernova feedback \citep{springel00_fb}
are treated using simplified recipes that were calibrated to match 
observed star-formation rates.

The structure of the fiducial progenitor disc galaxies mimics 
Sab-Sb spirals today, similar to the Milky Way.
The dark-matter halo virial mass is $116\times 10^{10}\msun$,
represented by 120,000 particles (of $0.97\times 10^{7}\msun$ each).
The virial radius is $272$ kpc, corresponding to a virial velocity 
of $135\kms$.
The halo density profile is NFW \citep{nfw97}
with a concentration parameter $C = 6$,
and a spin parameter $\lambda = 0.05$.
In the fiducial case, the baryonic fraction of the total mass is $0.054$.
The total baryons are divided into a stellar disc fraction of 0.66,
a gaseous disc fraction of 0.20, and a stellar bulge fraction of 0.14.
The stellar disc mass of $4.11\times 10^{10}\msun$ is represented by 50,000
particles (of $0.822\times 10^{6}\msun$ each).
The surface density profile is exponential, with a scale radius $\Rd=2.85$ kpc,
and the stability is ensured by a Toomre $Q$ parameter of 2.  
% z0 = 0.125 Sigma0,disc = 0.0805351186861026
%
The fiducial gas disc of $1.22\times 10^{10}\msun$ is 
represented by 50,000 particles (of $2.44\times10^{5}\msun$ each).
The gas is distributed in an exponential disc of a scale radius $3 \Rd$
extending out to 5 gas scale radii.
The fiducial bulge is a sphere containing $0.89\times 10^{10}\msun$ of stars
in 20,000 particles. Its density profile is exponential with scale radius
$0.62$ kpc.
The smoothing length is $h=100$ pc for the gas and star particles
and $400$ pc for the dark-matter particles,  
with the force becoming Newtonian at $\geq 2.3h$.
%Tully-Fisher Vflat = 193.188029846181 km/sec (from Bell & de Jong 2001)
%Baryonic specific angular momentum jd= 0.0267933

The two galaxies, which were set on a parabolic orbit, 
merge because of dynamical friction due to their massive haloes.
One galaxy starts such that its disc lies in the collision orbital plane,
and the other galaxy's disc is inclined by $30^\circ$, both with a prograde
orientation of spin.
The initial separation between the galaxy centres is $250$ kpc, 
and the expected pericentre (assuming point-like galaxies) is at $13.6$\kpc,
namely about 5\% of the virial radius.
The merger results in two successive starbursts, one after the first close 
approach, and the other after the second, final coalescence.  The starbursts
occur between 2 and 3 Gyr after the beginning of the simulation, and the
remnant is ``observed" at 4 Gyrs. 
The amount of stars formed during the merger is roughly proportional to the
initial gas fraction, and is not too sensitive to the  orbit or orientation.  
In the cases with moderate gas fractions,
the young stars formed during the merger typically constitute 
$\sim 20\%$ of the total stars.
 
The remnant galaxies resemble normal elliptical galaxies in many respects,
as demonstrated in \citet{dekel05}. This includes the stellar density profile,
the ellipticity and triaxiality, the velocity-dispersion profile as well 
as higher moments of the line-of-sight velocity distribution.

The galaxies in all cases are identical in their dark matter halos,
baryonic fraction and merger parameters.
In one suite of simulations the bulge-to-disc ratio is the fiducial 0.17,
while in the other suite it is 0.39.
Each of the two suites consists of 5 cases, 
where the bulge-to-disc ratio is kept fixed and {\it only\,} the disc
gas-to-stars mass ratio $g$ is varied.
In the small-bulge suite, $g$ varies between 0.12 and 3.0 
[corresponding to gas/(gas+stars) between 0.11 and 0.75],
while in the suite with a more massive bulge, $g$  
varies between 0.097 and 1.72 [corresponding to gas/(gas+stars) between 0.088 
and 0.63].

%The fraction of the gas which formed stars during the merger
%was found to be 0.69, 0.71, 0.74 respectively,
%namely a stellar-mass growth factor of $\Le/\Ld=$ 1.17, 1.51, 2.02.
%The effective radius of the remnant is $\Re=$ 3.8, 2.2, 1.6 $\kpc$.
%The effective velocity dispersion is $\sigma_e=$ 181, 203, 224 $\kms$.
%The growth factors scale like these quantities because the initial
%discs were identical in all three cases.
%OLD:
%[When addressing the same growth factors as functions of gas fraction 
%$=g/(1+g)$ instead of $g$, the power-law fits yield 
%$\mm'=0.51$, $\vv'=0.20$, $\rr'=-0.81$,
%but the small deviations from power laws are slightly larger.]

\begin{figure}
\vskip 9.8cm
{\includegraphics{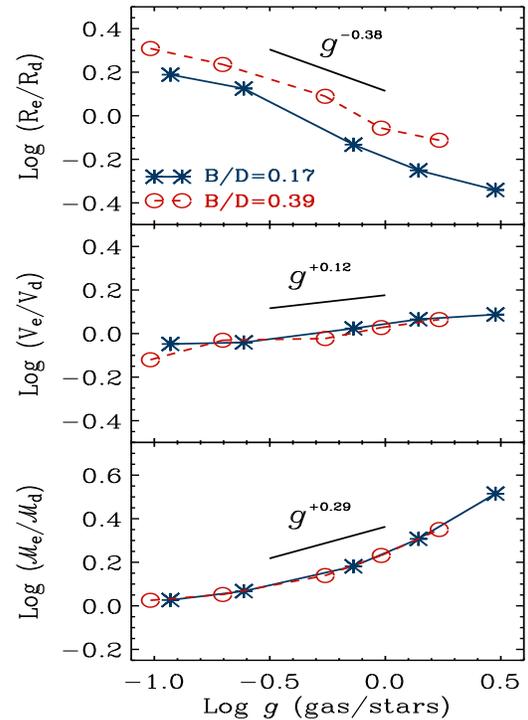}} % fig1_tj.eps
\caption{Dependence on gas content in merger simulations.
The relative changes in effective radius $R$, internal velocity $V$ 
and stellar mass $\ms$ are plotted against the initial gas-to-star ratio $g$
The two curves, which refer to different bulge fractions, show similar slopes.
The effective power-law fits are indicated.
}
\label{fig:fp_gp}
\end{figure}

\Fig{fp_gp} shows the growth factors in the mergers as a function of $g$.
The trends are fit by the power laws in \equ{growth} 
with $\rr \simeq -0.4$, $\vv \simeq 0.12$, $\mm \simeq 0.3$.
The radius growth factor shows the strongest dependence on $g$.
The power laws are good approximations for the $g$-dependence of the 
radius and velocity factors over the whole range of $g$,
but the power-law fit in the case of stellar mass is rather crude.  
The effective slope of $\mm \simeq 0.3$ coincides with the local slope near 
$g \simeq 0.5$, and it can serve as a sensible approximation over a $g$-range
of somewhat less than a decade about it. 
At very low $g$ values the slope is as small as $\sim 0.1$, and at very high
$g$ values it becomes steeper than $0.5$.   
Still, the crude power-law fit can adequately serve our simplified modeling
of the tilts.

While the magnitude of the radius shrinkage depends on the bulge-to-disc
ratio, we find that the power-law fits are insensitive to the presence
of a more massive bulge. This indicates that our results should be robust 
to the exact shape of the progenitors.

When the growth factors are plotted instead as functions of 
$g'=$gas/(gas+stars) rather than $g$, the corresponding best-fit
power indices become $\rr' \simeq -0.6$, $\vv' \simeq 0.2$, $\mm' \simeq 0.4$. 
However, the power-law fits in this case are not as good. 
In particular, the $R$ factor flattens at the low-$g'$ end, 
and the $\mm$ factor steepens significantly at the high-$g'$ end.
We therefore prefer to use $g$ as the gas-content variable.

%5
%%%%%%%%%%%%%%%%%%%%%%%%%%%%%%%%%%%%%%%%%%%%%%%%%%%%
\section{Results: Progenitor Properties}
\label{sec:results}

\begin{figure}
\vskip 8.1cm
{\includegraphics{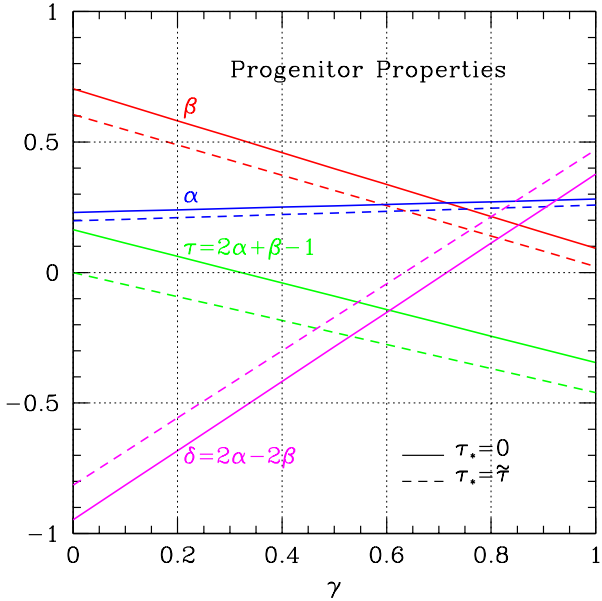}} %g.eps
\caption{Progenitor properties (the subscript ``d" is omitted)
that lead by mergers to the observed structural indices of ellipticals,
$\talphae=0.23$ and $\tbetae=0.70$ ($\ttaue=0.16$).
The values of the parameters along each vertical line correspond 
to a solution of \equs{tau}-(\ref{eq:r}).
The power index $\gamma$ refers to the decline of gas-fraction with mass. 
The indices $\alpha$ and $\beta$ are the slopes of the $V\ms$ and $R\ms$
scaling relations.
The tilt $\tau$ refers to variations in the ratio of mass to stellar mass
within the effective radius.
The index $\delta$ refers to variations in mean total density within 
the effective radius.
The solid and dashed curves correspond to $\taus=0$ and $\taus=\ttaue$,
where the tilt of the FP compared to the virial plane is either all structural
or all due to stellar populations respectively.}
\label{fig:g}
\end{figure}

We are now set to solve \equs{tau}-(\ref{eq:r}).
We use the values of the growth-factor indices $\rr=-0.4$, $\vv=0.12$ and 
$\mm=0.3$ as deduced from the merger simulations.
We then adopt the observed values for ellipticals
$\talphae=0.23$ and $\tbetae=0.70$ ($\ttaue=0.16$) 
and solve for the required progenitor properties.
The one-parameter family of solutions is shown in \fig{g}. 
For each choice of $\gamd$, 
the associated values of the power indices $\alphad$ and $\betad$,
and their linear combinations $\taud$ and $\deld$, 
can be read from the corresponding curves.
For example,
the vertical line at $\gamd=0$ refers to the null case where the progenitors 
are identical to the final ellipticals because the available
gas fraction does not vary with mass.
Any other vertical line corresponds to another possible solution for
the progenitor properties.
Higher initial values of $\gamd$ naturally correspond to
larger changes in the power indices as a result of mergers.
 
We see that
the slope $\alpha$ of the $V\ms$ relation does not change much by 
dissipative mergers even if $\gamd$ is large.
This is because both $V$ and $\ms$ increase during the merger in a way that
makes the systems propagate roughly along the $V\ms$ relation.
Also, given that $\vv=0.12$ is so small, and that both $\gamma$ and $\mm$ are
most likely smaller than unity, \equ{v} says that $\alphad$ is just a bit
larger than $\alphae$, quite insensitive to the actual value of $\gamd$.
The observed slope of $\talphae \simeq 0.23$ thus puts a strong robust 
constraint on the corresponding progenitor slope, forcing it to be
well within the range $0.20$-$0.28$.
This rules out in particular progenitors that lie on the naive 
virial-spherical plane $\alphad=\betad=1/3$ (but we will see below that
this conclusion is limited to the scaling relations based on stellar mass,
and becomes invalid when all the baryonic mass is considered).

On the other hand, the change in the slope $\beta$ of the $R\ms$ relation 
is strongly dependent on $\gamd$ and can be quite large.
This is because the radius shrinkage is the 
dominant change, helped by the fact that $\rr$ and $\mm$ in \equ{r} 
are of opposite signs.  Correspondingly, the values of $\tau$ and $\delta$ 
suffer changes which are strong functions of $\gamd$: the former grows
and the latter declines as a result of mergers.

If we limit the discussion to conventional models of galaxy formation
which do not permit an increasing density with progenitor mass, 
the corresponding requirement $\deld \leq 0$ implies   
an upper limit of $\gamd \leq 0.67$ (for $\taus = 0.1$).
This corresponds to $\betad \geq 0.25$ and $\taud \geq -0.25$.

If we also require that the initial $M/\ms$ is not increasing with
progenitor mass, based for example on the common wisdom that supernova 
feedback becomes less effective
in massive haloes \citep{ds86}, then the requirement $\taud \leq 0$
puts a lower bound of $\gamd \geq 0.15$ ((for $\taus = 0.1$).
This corresponds to $\betad \leq 0.55$ and $\deld \geq -0.7$.

Progenitors with $\gamd \simeq 0.4-0.6$ are favorable not only because
these values are compatible with today's trend of gas fraction along the
blue sequence, but also because they represent solutions which are
close to homology, roughly obeying the virial theorem ($\taud \lsim 0$) 
and a weak decline of density with mass ($\deld \lsim 0$).
For example, 
a typical value of $\gamma = 0.55$, compatible with today's gas gradient
along the blue sequence plus an estimated contribution from the 
mass dependence of the cooling efficiency, corresponds to
$\alphad \simeq =0.25$, $\betad \simeq 0.31$, $\taud \simeq -0.20$
and $\deld \simeq -0.15$.
 
% \taus
We learn from \fig{g} that the value of $\taus$, namely the way the stellar
$\ms/L$ varies with mass, has a surprisingly small effect on the required 
progenitors. Even a choice of a large stellar tilt, $\taus = \ttaue =0.16$, 
which eliminates the need for a structural tilt of the FP (namely $\taue=0$), 
still implies strong changes in $\beta$ (and $\tau$ and $\delta$)
between the progenitors and the remnants
once $\gamd$ is not negligible. 

%6
%%%%%%%%%%%%%%%%%%%%%%%
\section{Interpreting the Progenitors}
\label{sec:interpret}

The solutions obtained above can be compared to what might be
expected based on other considerations for the properties of the 
early gaseous progenitors of ellipticals.

% virial
Recall that the naive model of top-hat collapse to virial equilibrium 
followed by disc formation conserving angular momentum, at the same epoch 
for all galaxies, yields 
$\alphad=\betad=1/3$, namely progenitors that lie on the virial plane 
($\taud=0$) with the same effective density for galaxies of all masses 
($\deld=0$).
As noticed in \fig{g}, this is not a proper solution.
A solution with $\betad=\alphad$ does exist, for $\gamd \simeq 0.67$, 
but it has the ``wrong" value of $\betad=\alphad \simeq 0.25$. 
A value close to $\alphad \simeq 1/3$ cannot be obtained for any sensible
value of $\gamd$ as long as the observed slope is $\alphae \leq 0.23$,
because $\alpha$ hardly changes in the course of a merger. 

% C    V-M
However, a somewhat more sophisticated conventional model of disc 
formation in $\Lambda$CDM haloes does predict structural power indices in 
the range required for the progenitors of ellipticals. 
Studies of the profiles of dark haloes in 
cosmological simulations \citep{bullock01_c},
based on the NFW \citep{nfw97} functional form 
(but without limiting the generality of the
analysis), yield a mean trend of halo concentration with virial mass,
$\cv \prop \Mv^{-0.13}$,
reflecting the different effective formation
epoch as a function of halo mass \citep{bullock01_c,wechsler02}.
While the flat part of the circular velocity resembles the virial velocity,
$\Vv \prop \Mv^{1/3}$, 
the maximum NFW circular velocity is 
$\Vm \prop \cv^{0.27}\Vv \prop \Mv^{0.30}$.
If the relevant velocity characterizing the progenitors is between
$\Vv$ and $\Vm$,   
the model predicts a slope for the $\Vd$-$\msd$ relation in the range
$0.30 \leq \alphad/(1+\tauv) \leq 1/3$,
where $\tauv$ characterizes possible variations of virial to stellar 
mass ratio, $\Mv/\ms \prop \ms^\tauv$.

% C   R-M
The effective disc radius is evaluated by \citet{bullock01_j} based on the
dark-halo properties under the assumptions of a constant spin parameter 
as a function of mass (as seen in the simulations) and
conservation of angular momentum during the gas contraction.
For a disc baryonic fraction that is independent of mass,
and in the limit $\cv \gg 1$, the disc radius can be approximated by  
$R \prop \cv^{-0.7}\Rv \prop \Mv^{0.42}$. 
An increasing baryonic fraction with mass
would tend to counter-balance the $\cv^{-0.7}$ dependence on concentration,
so the model predicts a slope for the $\Rd$-$\msd$ relation in the range 
$1/3 \leq \betad/(1+\tauv) \leq 0.42$.

% solution compatible with model
This crude model based on the $\Lambda$CDM halo properties is compatible 
with solutions shown in \fig{g} for the required properties of
the progenitors of ellipticals.
If we take the model predictions to be $\alphad \simeq 0.32(1+\tauv)$ and
$\betad \simeq 0.37(1+\tauv)$, and assume as before $\taus \simeq 0.1$,
then the proper solution of \equs{r} and \ref{eq:v} is 
$\tauv \simeq -0.24$, $\gamma \simeq 0.6$,
$\alphad=0.24$ and $\betad=0.28$.
The corresponding tilt of $M/\ms$ is negative, $\taud \simeq -0.23$,
and the density decline is rather weak, $\deld \simeq -0.08$.

% gamma
The value of $\gamma \simeq 0.6$ is roughly compatible with today's gas 
gradient along the blue sequence of $\gamma \simeq 0.4$ \cite{kannappan04}. 
The difference can be attributed to the mass dependence of the 
cooling efficiency, when $g$ represents the gas that actually manages to 
cool on a dynamical time scale after the merger. This contribution to $\gamma$
can be crudely estimated from the 
steepening from $\betad \simeq 0.52$ to $\betae \simeq 0.44$
seen in Fig.~11 of \citet{robertson06} for mergers with a fixed gas fraction
$g'=0.4$ (corresponding to $g=0.67$). When inserted in \equ{r}, with
$\rr=-0.4$ and $\mm=0.3$, we obtain a crude estimate of
$\gamma \simeq 0.15$ for the cooling-efficiency contribution at a fixed $g$. 
A value of $\gamma \sim 0.5-0.6$ for the gaseous progenitors 
is therefore not surprising.
%D [add a comment about future work?]

% tauv taud
Unlike the final ellipticals, the gaseous progenitors are predicted to be
self-similar in the sense that the ratio of total-to-stellar mass scales
with mass in a similar way inside
the virial radius and inside the effective radius,
$\tauv \simeq \taud$.
The fact that this tilt is negative, $\simeq -0.24$ is not surprising either.
Such a tilt is seen below $\simeq 3\times 10^{10}\msun$
\citep{bell03_baryon,yang03}.
A negative tilt as steep as $\tauv \sim -0.4$ is deduced from the
``fundamental line" of dwarf galaxies and predicted
by crude energy considerations of supernova feedback in the limit where
a small fraction of the gas has turned into stars
very effective feedback \citep{dw03,db06}.

% beta SDSS
The slope $\betad \simeq 0.28$ lies well inside the range of the
slopes observed for late-type galaxies in the SDSS:
$\betad \simeq 0.15$ and $0.40$ below and above 
$\ms \simeq 4\times 10^{10}\msun$ respectively \citep{shen03,kauf03_pop}.
A similar slope of $\betad \simeq 0.29$ is also deduced from a study of
a large local sample of late-type spirals, adopting $\ms/L_I \prop L_I^{0.15}$
for these galaxies \citep{cour06}.
The $R$-$\ms$ relation for typical progenitors thus resembles 
that of today's intermediate, $\sim L_*$ spirals.

% alphad
The value of $\alphad \simeq 0.24$ is compatible with the slope
of the ``baryonic" Tully-Fisher relation by \citet{mcgaugh05},
referring to the flat part of the rotation curve.
It is perhaps a bit lower than the $\alphad \simeq 0.28$
deduced using $\ms/L_I \prop L_I^{0.15}$ from \citet{cour06}, in which the 
rotation velocity is measured at 2.2 exponential disc radii.

%%%%%%%%%%%%%%%%%%%%%%%%%%
\section{Baryonic Relations}
\label{sec:baryonic}

The analysis above could be performed alternatively with 
the stellar mass $\ms$ replaced by the baryonic mass $\mb$ (of stars plus gas).
We use the same notation as above with a prime added to each variable.
This analysis is simplified by the fact that the change in $\mb$ during the
merger can be neglected, namely $\mm'=0$ in the analog of \equ{growth}. 
It also eliminates the need to deal with the poor power-law fit of
$\mse/\msd$ as a function of $g$ (\fig{fp_gp}).
The analogs of \equ{v} and \equ{r} become the explicit expressions
\be
\alphad'=\alphae'+\vv'\gamma' \ ,
\quad
\betad'=\betae'+\rr'\gamma' \ ,
\label{eq:vr_prime}
\ee
in which $\gamma'$ refers to the gas-to-baryon ratio in the progenitors, 
\be
g'\prop \mb^{-\gamma'} \ .
\ee

\begin{figure}
\vskip 7.4cm
{\includegraphics{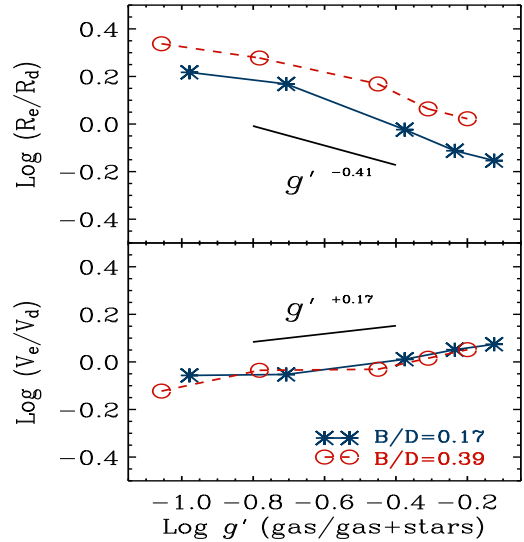}} % fig1_tj.eps
\caption{Same as \fig{fp_gp} but for the baryons (gas+stars) replacing the
stars.  The baryonic mass is preserved during the mergers.
}
\label{fig:fp_gp_prime}
\end{figure}

\Fig{fp_gp_prime} is the analog of \fig{fp_gp}, now as a function of the gas
fraction $g'$, showing reasonable power-law fits with slopes $\vv'=0.17$ and 
$\rr'=0.42$.

\begin{figure}
\vskip 8.1cm
{\includegraphics{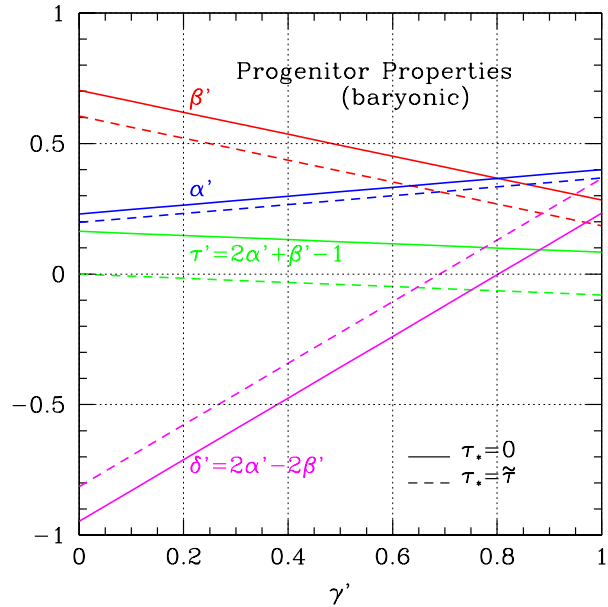}} %g.eps
\caption{Same as \fig{g} but for the baryons (gas+stars) replacing the
stars, showing the solutions to \equs{vr_prime}.
}
\label{fig:g_prime}
\end{figure}

The solutions of \equs{vr_prime} are shown in \fig{g_prime}.
With $\mm'=0$, the changes in $\alpha'$ and $\delta'$ as a function of
$\gamma'$ are stronger than they were in $\alpha$ and $\delta$ as a function
of $\gamma$,
while the changes in $\beta'$ and $\tau'$ are weaker than before.
We see in particular that a proper solution is
$\alphad' \simeq \betad' \simeq 0.33$,
namely $\taud' \simeq \deld' \simeq 0$,
with $\gamma' \simeq 0.7$.
This is the only solution for which neither the effective
total-to-baryonic mass nor the effective density are increasing with mass. 
If the baryonic mass scales with the virial mass,
these progenitors are the homologous family
predicted by the simplest model of spherical collapse into virial equilibrium.
This indicates no significant loss of baryons from the progenitors
due to feedback or other effects. The feedback may have an important
role in determining the strong variation of gas-to-baryon ratio with 
progenitor mass.

%%%%%%%%%%%%%%%%%%%%%%%%%%%%%%
\section{Conclusion}
\label{sec:conc}

% dissipation gradient is the key
We have clarified how the deviations of the structural scaling relations
of elliptical galaxies from the expectations based on the virial theorem
and an assumed homology of the elliptical family
could arise from the dissipative processes in major mergers.
The key is a systematic variation of the gas fraction available for dissipation
as a function of progenitor mass.
The tilt of the elliptical's fundamental plane compared to the virial plane
and the steep decline of density with mass are driven by the
differential shrinkage of the effective stellar radius
as a function of the degree of dissipation involved in the merger.

% alpha change little
The correlated enhancements in internal velocity and stellar mass during a
merger, and the fact that the change in velocity is rather small, lead
to only a small change ($\sim 10\%$) in the log-slope of the 
velocity-stellar mass relation, keeping it near $V\prop \ms^{1/4}$ for 
both the merger progenitors and products. This is in the ball-park of
the stellar Tully-Fisher relation observed in today's disc galaxies
\citep{mcgaugh05,cour06}.
The small changes in velocity indicate that the tilts are primarily
associated with structure rather than kinematics.

% beta
The predicted log-slope of the radius-stellar mass relation in the progenitors
is roughly $R \prop \ms^{0.3}$, compatible with the observed relation for
intermediate late-type galaxies near 
$\ms \sim 10^{10-11}\msun$ \citep{shen03,cour06}. 
It is steeper than the slopes observed at the faint end ($\betad \simeq 0.15$) 
and at the bright end ($\betad \simeq 0.4$) of late-type galaxies. 

% big discs at high z
The similarity of the progenitors to today's relatively big discs may 
imply that the latest dissipative mergers in the history of 
ellipticals tended to involve grown discs rather then smaller dwarfs. 
This is compatible with the detections of big gaseous discs at $z\sim 2-4$
\citep{tacconi_genzel06}, which could be produced by cold flows \citep{db06}, 
and with the histories of big ellipticals in cosmological
simulations that properly match the galaxy bimodality at low and
high redshifts \citep[][Cattaneo, Dekel \& Faber, in preparation]{cattaneo06}.
However, it does not rule out later non-dissipative mergers along the 
red sequence, which tend to preserve the FP and the associated scaling 
relations.\footnote{This is yet to be tested using simulations
of several successive non-dissipative mergers between spheroids.} 
The tilts can be explained by dissipative major mergers, or related
dissipative processes, somewhere in the histories of elliptical galaxies.

% taud
The required progenitor properties match the expectations from a  
simple model of disc formation in $\Lambda$CDM haloes 
\citep{mmw98,bullock01_c}. 
First, the progenitors are predicted to lie on a plane that is tilted relative
to the virial plane in the opposite sense to the tilt of the fundamental
plane of ellipticals, roughly $M/\ms \prop \ms^{-(0.2-0.3)}$. 
This is similar to the expectations from the effect of supernova feedback. 
The tilt becomes positive as a result
of the dissipative shrinkage of the baryonic component in the mergers. 
 
% deltad
Second, The progenitors are expected to be close to homologous,
in the sense that the dark-matter fractions within the effective radius
and the virial radius scale similarly with mass.
They should have a very weak decline of density with mass,
roughly $\rho \prop \ms^{-0.1}$, which later steepens by the differential
dissipative shrinkage to near $\rho \prop \ms^{-0.9}$ in the merged 
ellipticals.

% gamma
Our main prediction is that the variation with progenitor mass of the 
dissipative gas-to-star ratio should be roughly 
$g \prop \ms^{-\gamma}$ with $\gamma \simeq 0.5$-$0.6$. 
This is consistent with the observed trend ($\gamma \simeq 0.4$) in today's 
blue-sequence galaxies \citep{kannappan04},
after subtracting a certain contribution to $\gamma$ from the dependence 
of cooling efficiency on mass at a fixed gas fraction.
In an ongoing study we attempt to disentangle in more detail between the 
roles of gas-fraction and cooling efficiency. This will be achieved by 
performing merger simulations similar to the ones described in this paper 
but between galaxies of higher and lower masses.

% baryonic
The corresponding baryonic relations, in which the stellar mass is replaced by
the total baryonic mass in gas and stars, are consistent with the simplest
homology and spherical virial equilibrium, $V \propto R \propto \mb^{1/3}$,
namely negligible variations in $M/\mb$ and in mean density.
While the baryonic fraction is hardly changing, the gas fraction is predicted
to vary strongly with baryonic mass in the progenitors, $g'\prop \mb^{-0.7}$.

% bulge
The little relevance of the bulge fraction to the slopes of the scaling
relations indicates that the gaseous progenitors need not necessarily be
pure discs, as long as they have a proper systematic decline of gas fraction 
with mass.

% taus xxx
Our qualitative conclusions are insensitive to the actual contribution
of variations in stellar mass-to-light ratio to the tilt.
A substantial variation in the stellar mass-to-light ratio, 
on the order of $\ms/L \prop \ms^{0.1}$,
could in principle contribute to the tilt of the fundamental plane,
but it cannot give rise to ёthe observed strong decline of density with mass.
This indicates that the structural changes due to differential dissipation
effects in mergers should have an important role in shaping up the properties
of ellipticals, and implies that the trend in available gas fraction
as a function of progenitor mass is a robust prediction.

% observed predictions
High-redshift observations should identify the predicted early
progenitors of today's ellipticals, namely relative massive, 
gas-rich discy galaxies with a significant decline of gas fraction
as a function of mass.  
Indeed, an analysis of the morphology of $z\sim 1$ galaxies
using a non-parametric classification system \citep{lotz06} suggests
that a large fraction of these galaxies are massive discs.  
Since this class of object makes up a significant fraction of the 
star formation, they are likely to be gas-rich.  
Observations at higher redshifts in the range $z \sim 2$-$4$,
both in the UV and sub-millimeter wavelengths,
have revealed gas rich galaxies of $\sim 10^{13}\lsun$, 
whose kinematics resemble in many cases rotating thick discs of 
radii $\sim 2$-$3$kpc \citep{tacconi_genzel06}. 
These galaxies may represent the required progenitors.
A straightforward test of the differential dissipation
model would be the detection of the predicted
systematic decline of gas fraction with mass in such high-redshift galaxies.

%%%%%%%%%%%%%
\section*{Acknowledgments}
We acknowledge stimulating discussions with M.~Covington,
S.M.~Faber, G.~Novak, and J.R.~Primack.
This research has been supported by ISF 213/02 and NASA ATP NAG5-8218.
AD acknowledges support from
a Miller Visiting Professorship at UC Berkeley,
a Visiting Professorship at UC Santa Cruz, and a Blaise Pascal International
Chair by Ecole Normale Superiere at the Institut d'Astrophysique, Paris.

%%%%%%%%%%%%%%%%%%%%%%%%%%%%%%%%%%%%%%%%%%%%%
\bibliographystyle{mn2e}
\bibliography{dekel}
%%%%%%%%%%%%%%%%%%%%%%%%%%%%%%%%%%%%%%%%%%%%%

\label{lastpage}
\end{document}